\documentclass[aps,prl,twocolumn,showpacs]{revtex4-1}

\usepackage{graphicx}
\usepackage{latexsym}
\usepackage{amsmath,amscd}
\usepackage{times}
\usepackage{epstopdf}

\begin{document}

\title{\bf\noindent Number of Common Sites Visited by $N$ Random Walkers}
\author{Satya N. Majumdar$^1$ and Mikhail V. Tamm$^2$}
\affiliation{ $^1$ Laboratoire de Physique Th\'eorique et Mod\`eles Statistiques (UMR 8626 du CNRS),
Universit\'e Paris-Sud, B\^at.\ 100, 91405 Orsay Cedex, France\\
$^2$ Physics Department, Moscow State University, 119992, Vorobyevy Gory, Moscow, Russia}

\pacs{05.40.Fb, 05.40.Jc, 02.50.Cw, 24.60.-k}

\begin{abstract}

We compute analytically the mean number of {\em common} sites, $W_N(t)$, visited by $N$ independent random
walkers each of length $t$ and all starting at the origin at $t=0$ in $d$ dimensions. We show that in the
$(N-d)$ plane, there are three distinct regimes for the asymptotic large $t$ growth of $W_N(t)$.  These three
regimes are separated by two critical lines $d=2$ and $d=d_c(N)=2N/(N-1)$ in the $(N-d)$ plane. For $d<2$,
$W_N(t)\sim t^{d/2}$ for large $t$ (the $N$ dependence is only in the prefactor). For $2<d<d_c(N)$,
$W_N(t)\sim t^{\nu}$ where the exponent $\nu= N-d(N-1)/2$ varies with $N$ and $d$. For $d>d_c(N)$, $W_N(t)\to
{\rm const.}$ as $t\to \infty$. Exactly at the critical dimensions there are logaritmic corrections: for
$d=2$, we get $W_N(t)\sim t/[\ln t]^N$, while for $d=d_c(N)$, $W_N(t)\sim \ln t$ for large $t$. Our
analytical predictions are verified in numerical simulations.

\end{abstract}
\maketitle

Computing the average number of distinct sites visited by a {\em single} $t$-step random walker
on a $d$-dimensional lattice, denoted by $S_1(t)$, is by now a classic problem with a variety of
applications ranging from the annealing of defects in crystals to the size of the territory
covered by a diffusing animal during the foraging period. First posed and studied by Dvoretzky
and Erd\"os in 1951~\cite{DE}, this problem has been solved exactly in a number of papers in the
1960's~\cite{Vineyard,MW}. It is well established (see ~\cite{Hughes} for a review) that
asymptotically for large $t$, $S_1(t)\sim t^{d/2}$ for $d<2$, $\sim t/\ln(t)$ for $d=2$ and
$\sim t$ for $d>2$. These results have been widely used in a number of applications in
physics~\cite{HK,HB,BG}, chemistry~\cite{Rice}, metallurgy~\cite{BD,Beeler,Rosenstock}, and
ecology~\cite{Pielou,E-K}. In 1992, Larralde and coworkers generalized this problem to the case
of $N$ independent random walkers (each of $t$ steps) all starting at the origin of a
$d$-dimensional lattice~\cite{Larralde}. They computed analytically $S_N(t)$, the mean number of
sites visited by {\em at least} one of the $N$ walkers in $d$ dimensions and found two
interesting time scales associated with the growth of $S_N(t)$. In the ecological context,
$S_N(t)$ represents the mean size of the territory covered by an animal population of size $N$.
The original results of Larralde et. al. have subsequently been corrected~\cite{YA}, used and
generalised in a number of other
applications~\cite{Havlin,Shlesinger,LWS,MMB,YL,Yuste,BH,Drager,Acedo,LW,LBW}.

\begin{figure}
\includegraphics[width=1.1\hsize]{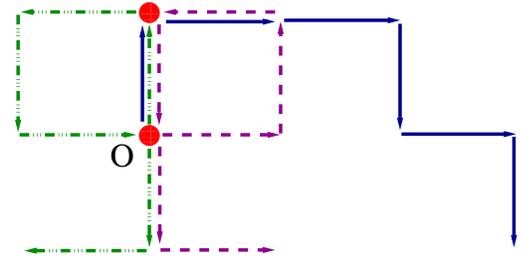}
\caption{(color online) A
realization of $3$ random walks, each of $6$ steps (denoted respectively by solid (blue), dashed
(purple), and dashed-dotted (green) lines) on a square lattice, all starting at the origin $O$.
There are two sites (each marked by a filled (red) circle) that are visited by all $3$ walkers.}
\label{fig_common}
\end{figure}

\begin{figure}
\includegraphics[width=.9\hsize]{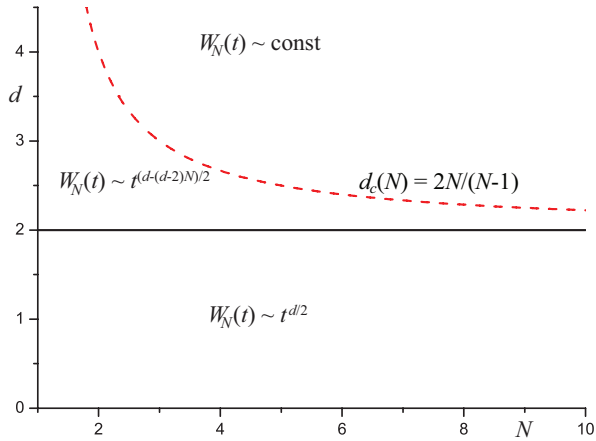}
\caption{(color online) In the $(N-d)$ plane, there are two critical
lines $d=2$ (lower horizontal line) and
$d_c(N)=2N/(N-1)$ (upper dashed (red) curve). The mean number of common sites
$W_N(t)$ visited by $N$
walkers, each of $t$ steps and all starting at the origin at $t=0$, have different
asymptotic behaviors for
large $t$ in the three regimes $d<2$, $2<d<d_c(N)$ and $d>d_c(N)$.}
\label{fig_phd}
\end{figure}

In this Letter, we study a complementary question: what is the average number of {\em common} sites,
$W_N(t)$, visited by $N$ independent walkers, each of them consisting of $t$ steps and
starting at the origin
of a $d$-dimensional lattice? A typical realization in $d=2$ for $N=3$ walkers is shown in Fig.
(\ref{fig_common}). Our exact results demonstrate that $W_N(t)$ exhibits a rather
rich asymptotic behavior
for large $t$. In the $(N-d)$ plane ($N$ being the number of walkers,
or the population size in ecological
context, and $d$ -- the space dimension)
we find an interesting phase diagram where two critical lines $d=2$ and
$d_c(N)=2N/(N-1)$ separate three phases with different asymptotic growth
of $W_N(t)$ (see Fig.
(\ref{fig_phd})). For large $t$, we show that
\begin{eqnarray}
W_N(t) & \sim & t^{d/2}\quad {\rm for}\,\, d<2 \nonumber\\
&\sim & t^{\nu} \quad {\rm for}\,\, 2<d< d_c(N)=\frac{2N}{N-1} \nonumber \\
&\sim & {\rm const.} \quad {\rm for}\,\, d>d_c(N)
\label{wnt.asymp1}
\end{eqnarray}
where the exponent $\nu= N-d(N-1)/2$ varies with $N$ and $d$.
Exactly at the two critical dimension, there
are logarithmic corrections. In particular, for large $t$,
$W_N(t)\sim t/[\ln t]^N$ in $d=2$, and $W_N(t)\sim
\ln t $ in $d=d_c(N)$ (with $N>1$).
The existence of the intermediate phase $2<d<d_c(N)=2N/(N-1)$, with a
growth exponent $\nu$ varying with $N$ and $d$, is perhaps the most striking of our
results. For instance, for $N=2$ we have $d_c(2)=4$ and so in $2<d=3<4$ our result predicts $\nu=1/2$, i.e.,
$W_2(t)\sim t^{1/2}$, a prediction that is verified in our numerical simulations.

The statistics of the number of most popular sites, i.e., the sites visited by all the walkers
arises quite naturally in a number of contexts such as sociology, ecology, artificial
networks (e.g., internet, transport and engineering networks) and polymer networks
just to name a few.
For example in a multiple user network such as
the internet, the most popular `hub' sites visited by all the users are known to play very
important role in the dissemination of information~\cite{NetworkBook}. The knowledge of how many
of them are there is fundamental for many applications. In tourism industry, it is important to
know the number of most popular sites in a given area or city that are visited by all the
tourists.
Motivated by this general question, in this Letter we study the statistics of the number of most
popular sites in perhaps the simplest model, namely for $N$ independent random walkers in the
$d$-dimensional space and show that even in this simple model, the asymptotic temporal growth of
the mean number of common sites frequented by all $N$ walkers exhibits surprisingly rich
behavior. We show that our results also have close connections to the probability of
non-intersection of random walks studied in the mathematics literature~\cite{Lawler,Duplantier}.
Given the abundance of random walks used as a fundamental model to study numerous natural and
artificial systems, and the richness of our exact results, we believe that they will be useful
in more specific applications in the future.

We consider $N$ independent $t$-step walkers on a $d$-dimensional lattice, each starting at the origin. To
compute the number of common sites visited by all the $N$ walkers, it is first useful to introduce a binary
random variable $\sigma_{k,N}(\vec x,t)$ associated with each site $\vec x$ such that $\sigma_{k,N}(\vec
x,t)=1$ if the site $\vec x$ is visited by exactly $k$ of the $N$ walkers and $\sigma_{k,N}(\vec x,t)=0$
otherwise. Then the sum $V_{k,N}(t)= \sum_{\vec x} \sigma_{k,N}(\vec x,t)$ represents the number of sites
visited by exactly $k$ of the $N$ walkers each of $t$ steps in a particular realization of the walks.
Clearly, $V_{k,N}(t)$ is a random variable that fluctuates from one sample to another. Taking average gives
the mean number of sites visited by exactly $k$ walkers, $\langle V_{k,N}(t)\rangle = \sum_{\vec x}
P_{k,N}(\vec x,t)$ where $P_{k,N}(\vec x,t)=\langle \sigma_{k,N}(\vec x,t)\rangle $ is the probability that
the site $\vec x$ is visited by exactly $k$ of the $N$ walkers. Since the walkers are independent, one can
write
\begin{equation}
P_{k,N}(\vec x,t)= \binom{N}{k}\, \left[p(\vec x,t)\right]^k\,\left[1-p(\vec x,t)\right]^{N-k}
\label{binom.1}
\end{equation}
where $p(\vec x,t)$ is the probability that the site $\vec x$ is visited by a {\em single} $t$-step walker
starting at the origin. Thus
\begin{equation}
\langle V_{k,N}(t)\rangle= \binom{N}{k}\, \sum_{\vec x} \left[p(\vec x,t)\right]^k\,\left[1-p(\vec
x,t)\right]^{N-k}\,.
\label{binom.2}
\end{equation}
Finally, the mean number of {\em common} sites visited by all the $N$ walkers is simply
\begin{equation}
W_N(t)= \langle V_{N,N}(t)\rangle= \sum_{\vec x}\left[p(\vec x,t)\right]^N\,.
\label{common.1}
\end{equation}
Hence, once the basic quantity $p(\vec x,t)$ for a single walker is known, we can determine $\langle
V_{k,N}(t)\rangle$ and in particular $W_N(t)$ just by summing over all sites as in Eq. (\ref{common.1}). Note
that, by definition, $p(0,t)=1$ for all $t\ge 0$ since the walker starts at the origin.

The probability $p(\vec x,t)$ can be fully determined for a lattice walker with discrete time steps using the
standard generating function technique~\cite{MW}. However, since we are interested here mainly in the
asymptotic large $t$ regime, it is much easier to work directly in the continuum limit where we treat both
space $\vec x$ and time $t$ as continuous variables. Consider then a single Brownian motion of length $t$ and
diffusion constant $D$ in $d$-dimensions starting at the origin. We are interested in $p(\vec x, t)$, the
probability that the site $\vec x$ is visited (at least once) by the walker up to time $t$. Let $\tau$ denote
the last time before $t$ that the site $\vec x$ was visited by the walker. Then, clearly
\begin{equation}
p(\vec x,t)= \int_0^t G(\vec x,\tau)\, q(t-\tau)\, d\tau
\label{convol.1}
\end{equation}
where $G(\vec x,\tau)= e^{-x^2/{4Dt}}/(4\pi\, D\,t)^{d/2}$ (where $x=|\vec x|$) is the standard Green's
function denoting the probability that the particle is at $\vec x$ at time $\tau$ and $q(\tau)$ denotes the
persistence, i.e., the probability that starting at $\vec x$, the walker does not return to its starting
point up to time $\tau$. Note that $q(\tau)$ does not depend on the starting point $\vec x$ and is the same
as the probability of no return to the origin up to time $\tau$. Indeed, $q(\tau)= \int_{\tau}^{\infty}
f(\tau')d\tau'$ where $f(\tau)= -dq/d\tau$ is the standard first-passage probability to the
origin~\cite{Redner}.

The no-return probability $q(\tau)$ for a Brownian walker has been studied extensively and it is well known
that for large $\tau$, $q(\tau)\sim \tau^{d/2-1}$ for $d<2$, $q(\tau)\sim 1/{\ln \tau}$ for $d=2$, while it
approaches a constant for $d>2$ since the walker can escape to infinity with a finite probability for
$d>2$~\cite{Redner}. One can show that to analyze the large $t$ behavior of $p(\vec x,t)$ in Eq.
(\ref{convol.1}) in the scaling regime where $x\to \infty$, $t\to \infty$ but keeping $x/\sqrt{t}$ fixed, it
suffices to substitute only the asymptotic behavior of $q(\tau)$ in Eq. (\ref{convol.1}). This gives, for
large $t$
\begin{eqnarray}
p(\vec x,t) &\sim & \int_0^t G(\vec x,\tau) (t-\tau)^{d/2-1}\, d\tau\,\, {\rm for}\,\, d<2
\label{lowd}\\
p(\vec x,t) & \sim & \int_0^{t} G(\vec x, \tau) d\tau\,\, {\rm for}\,\, d>2
\label{highd}
\end{eqnarray}
where we have dropped unimportant constants for convenience. For $d=2$, $p(\vec x, t)\sim \int_0^t G(\vec x,
\tau) d\tau/{\ln (t-\tau)}$. Substituting the exact Green's function $G(\vec
x,\tau)=e^{-x^2/{4D\tau}}/(4\pi\, D\,\tau)^{d/2}$ one finds that $p(\vec x,t)$ has the following asymptotic
scaling behavior
\begin{eqnarray}
p(\vec x,t) &\approx & f_{<}\left(\frac{x}{\sqrt{4Dt}}\right)\,\, {\rm for}\,\, d<2
\label{lowd.scaling1}\\
p(\vec x,t) &\approx & t^{1-d/2}\,f_{>}\left(\frac{x}{\sqrt{4Dt}}\right)\,\, {\rm for}\,\, d>2
\label{highd.scaling1}
\end{eqnarray}
where the scaling functions for $d<2$ and $d>2$ can be expressed explicitly as
\begin{eqnarray}
f_{<}(z) &= & \int_0^1 e^{-z^2/u}\, u^{-d/2}\,(1-u)^{d/2-1}\, du \label{lowd.scaling2}\\
f_{>}(z) &= & \int_0^{1} e^{-z^2/u}\, u^{-d/2}\, du\, .
\label{highd.scaling2}
\end{eqnarray}
Exactly at $d=2$, one gets $p(\vec x,t) \approx (1/{\ln t}) f_2\left(x/\sqrt{4Dt}\right)$ where $f_2(z)=
\int_0^1 du\, e^{-z^2/u}/u$.

It is easy to derive the asymptotic tails of the scaling functions. One finds
\begin{eqnarray}
f_{<}(z) &\approx& {\rm const.} \quad {\rm as}\,\, z\to 0  \nonumber \\
&\approx & z^{-d}\, e^{-z^2} \,\, {\rm as}\,\, z\to \infty
\label{lowd.asymp1}
\end{eqnarray}
and
\begin{eqnarray}
f_{>}(z) &\approx & z^{-(d-2)} \quad {\rm as}\,\, z\to 0  \nonumber \\
&\approx & z^{-2}\, e^{-z^2} \,\, {\rm as}\,\, z\to \infty
\label{highd.asymp1}
\end{eqnarray}
At $d=2$, one finds $f_2(z)\sim -2 \ln (z)$ as $z\to 0$ and $f_2(z)\sim e^{-z^2}/z^2$ as $z\to
\infty$. Note that the scaling forms postulated in
Eqs. (\ref{lowd.scaling1}) and (\ref{highd.scaling1}) do not, in general, hold
for very small $x$.
For $d<2$, the scaling regime
can actually be extended all the way to $x\to 0$ and indeed, the exact relation $p(0,t)=1$
is actually part of the scaling regime. This is seen
by taking $x\to 0$
limit in Eq.
(\ref{lowd.scaling1}) and using the asymptotic small $z$ behavior of $f_{<}(z)$ in Eq.
(\ref{lowd.asymp1}). In contrast, for $d>2$, one can not recover $p(0,t)=1$ by taking $x\to 0$
limit in Eq. (\ref{highd.scaling1}). This is a manifestation of the fact that for $d>2$ one
always needs a finite lattice cut-off $a>0$ (see, e.g. ~\cite{Hughes}). Thus for $d>2$, the
continuum scaling result in Eq. (\ref{highd.scaling1}) does not hold for $x<a$.

We next substitute Eqs. (\ref{lowd.scaling1}) and (\ref{highd.scaling1}) in
Eq. (\ref{common.1}) and replace
the sum by an integral over space. Note that even though we started out with $d$ and $N$ being integers, the
general formula (\ref{common.1}) can be analytically continued to real $d>0$ and real $N>0$. So, from now on
we will consider $d$ and $N$ to be continuous real positive numbers as, e.g.,
represented in the phase
diagram in Fig. (\ref{fig_phd}). Indeed, non-integer values of $d$ can be interpreted in terms
of random
walks on fractal manifolds with non-integer dimensions.
Consider first the case $d<2$ where we get,
dropping unimportant prefactors, for large $t$
\begin{equation}
W_N(t)\sim t^{d/2} \int_0^{\infty} \left[f_{<}(z)\right]^N \, z^{d-1}\, dz\,.
\label{lowd.1}
\end{equation}
From the tails of the scaling function $f_{<}(z)$ in Eq. (\ref{lowd.asymp1}), it is evident that the integral
in Eq. (\ref{lowd.1}) is convergent and is just a constant and hence for $d<2$, $W_N(t)\sim b_N\, t^{d/2}$
for large $t$ with only the prefactor $b_N$, but not the exponent, depending on $N$. Exactly at $d=2$, using
$p(\vec x, t)\sim [1/\ln t]\, f_2(\left(x/\sqrt{4Dt}\right)$ and following a similar analysis we get for
large $t$
\begin{equation}
W_N(t) \sim \frac{t}{[\ln t]^N} \int_0^{\infty} [f_2(z)]^N\, z\, dz\,.
\label{d2.1}
\end{equation}
Using the exact form of the scaling function $f_2(z)$ described before, one can check that the integral above
is convergent and hence, for $d=2$, $W_N(t)\sim t/[\ln t]^N$ for large $t$.

For $d>2$, a similar manipulation is a bit more delicate. We recall that the scaling result for $p(\vec x,t)$
in Eq. (\ref{highd.scaling1}) holds only for $x>a$ where $a$ is a lattice cut-off, while $p(0,t)=1$
identically. Thus, in the sum in Eq. (\ref{common.1}) we separate $x=0$ term and replace the rest of the sum
by an integral over the scaling form
\begin{equation}
W_N(t)\approx 1 + A_d\, t^{N(1-d/2)}\,\int_{a}^{\infty} \left[f_{>}\left(\frac{x}{\sqrt{4Dt}}\right)\right]^N
\, x^{d-1}\, dx
\label{highd.0}
\end{equation}
where $A_d$ is a volume dependent constant and $a$ is the lattice cut-off. This gives, after rescaling
$z=x/\sqrt{4Dt}$
\begin{equation}
W_N(t)\approx 1+ A_d\, t^{N-(N-1)d/2}\,\int_{a/\sqrt{4Dt}}^{\infty} \left[f_{>}(z)\right]^{N}\, z^{d-1}\,
dz\,.
\label{highd.1}
\end{equation}
We now have to check how the integral behaves as $t\to \infty$, i.e., its lower limit approaches $0$. This is
controlled by the small $z$ behavior of the integrand. From Eq. (\ref{highd.asymp1}), we get
$[f_{>}(z)]^N\sim z^{-N(d-2)}$ as $z\to 0$. Hence the integrand behaves as $z^{d-(d-2)N-1}$ as $z\to 0$. Thus
two situations arise. If $d-(d-2)N>0$, i.e., $d< d_c(N)=2N/(N-1)$ (recall that $d>2$ already), the integral
is convergent at the lower limit and one can safely take the $t\to \infty$ limit and then Eq. (\ref{highd.1})
predicts that for large $t$ and $2<d<d_c(N)=2N/(N-1)$
\begin{equation}
W_N(t)\sim t^{\nu};\,\, \nu=N-d(N-1)/2\,.
\label{phase2.1}
\end{equation}
In contrast, if $d-(d-2)N<0$, i.e., $d> d_c(N)=2N/(N-1)$, the lower limit of the integral behaves as $\sim
t^{(N-1)d/2-N}$ for large $t$ which precisely cancels the power-law prefactor and
\begin{equation}
W_N(t)\to {\rm const.};\,\, d>d_c(N)=2N/(N-1)
\label{phase3.1}
\end{equation}
where the constant evidently depends on the cut-off, i.e., on the details of the lattice and is thus
nonuniversal. Physically this means that for $d>d_c(N)$, the common sites visited by all the walkers are
typically close to the origin and are visited at relatively early times. At late times, the walkers hardly
overlap and hence $W_N(t)$ does not grow with time. Finally, exactly at $d=d_c(N)$, a similar analysis shows
that $W_N(t)\sim \ln (t)$ for large $t$. The upper phase boundary in Fig. (\ref{fig_phd}) depicts the
critical line $d_c(N)= 2N/(N-1)$ as a function of $N$. Alternatively, for fixed $2<d<d_c(N)$, this critical
line can also be described as $N_c(d)= d/(d-2)$. For $1\le N\le N_c(d)$, we have $W_N(t)\sim t^{\nu}$ with
$\nu=N-d(N-1)/2$.

To check our analytical predictions, we have computed $W_N(t)$ numerically for $d=1$, $2$, $3$ and for
several values of $N$. In $d=1$, our result predicts that $W_N(t)\sim b_N t^{1/2}$ for large $t$ where the
exponent $1/2$ is independent of $N$ and only the prefactor $b_N$ depends on $N$. The results in Fig. (3a)
are consistent with this prediction. In $d=2$, our results predict that $W_N(t)\sim t/[\ln t]^N$ which is
verified numerically in Fig. (3b). For $d=3$, our result predicts that there is a critical value $N_c=3$ such
that $W_N(t)\sim t^{(3-N)/2}$ for $N<3$, $W_N(t)\sim \ln (t)$ for $N=3$ and $W_N(t)\sim {\rm const.}$ for
$N>3$. The simulation results for $d=3$ in Fig. (3c) are consistent with these predictions.

Interestingly, the critical dimension $d_c(N)=2N/(N-1)$ has also appeared in the probability
literature~\cite{Duplantier} in the context of the probability of no intersection of $N$ random walkers up to
$t$ steps all starting at the origin~\cite{Lawler}. To make a precise connection with our work presented
here, consider the random variable $V_{N,N}(t)$ that denotes the number of common sites visited by all the
$N$ walkers up to $t$ steps. Since all the walkers start at the origin, clearly the number of common sites
visited must be at least $1$ implying $V_{N,N}(t)\ge 1$. When $V_{N,N}(t)=1$, it corresponds to the event
that the walkers do not intersect further up to step $t$ and the origin at $t=0$ remains the only site
visited by all of them up to step $t$. Thus, the probability of no further intersection up to step $t$ is
$F_N(t)= {\rm Prob.} [V_{N,N}(t)=1]$. Lawler studied the decay of $F_N(t)$ for large $t$ rigorously in
special cases~\cite{Lawler} and Duplantier showed~\cite{Duplantier} that $F_N(t)$ approaches a constant as
$t\to \infty$ for $d>d_c(N)=2N/(N-1)$. For $d<d_c(N)$, $F_N(t)\sim t^{-\zeta}$ and the exponent $\zeta$ was
computed using an $\epsilon$ expansion around the critical dimension~\cite{Duplantier}. In contrast, in this
Letter we have computed the mean of the random variable $V_{N,N}(t)$, i.e., $W_N(t)=\langle
V_{N,N}(t)\rangle$. Note that while $F_N(t)$ is not exactly computable in all $d$, $W_N(t)$ is, as we have
shown here.

Another interesting related problem is to compute the mean number of $N$-fold self intersections of a single
ideal polymer chain of length $t$. In Ref. ~\cite{Khokh}, it was stated that in $d=3$ this grows as
$t^{(3-N)/2}$, which looks similar to our result $W_N(t)\sim t^{(3-N)/2}$ in the intermediate phase in $d=3$
and for $1<N< 3$. However, the two problems are not exactly identical and even the single chain result in
Ref.~\cite{Khokh} was qualitatively argued for, not rigorously proved, and the logarithmic correction for
$N=3$ was not mentioned.

\begin{widetext}

\begin{figure}
\includegraphics[width=1.0\hsize]{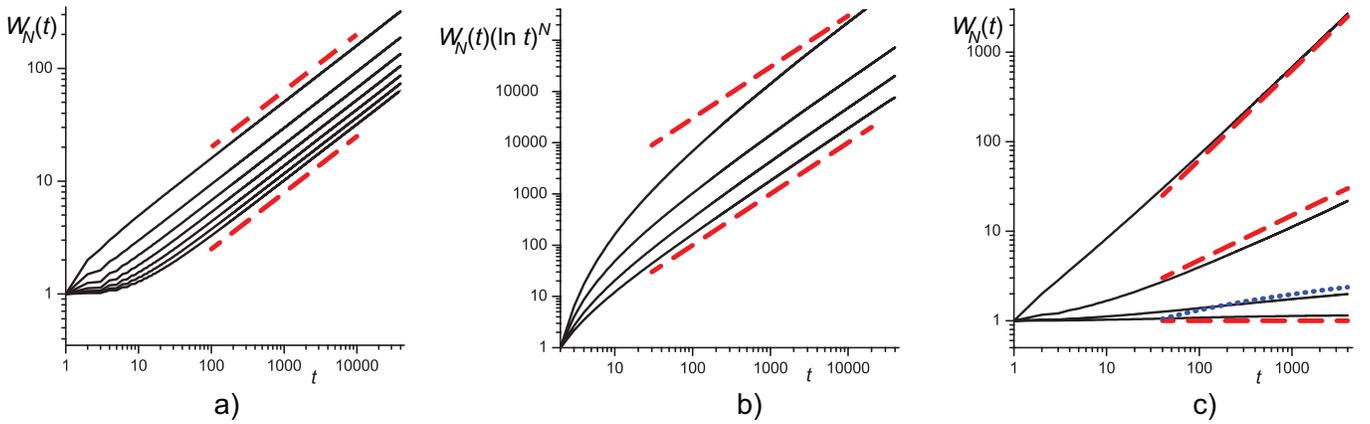}
\caption{ (color online) Numerical results for $W_N(t)$ vs. $t$ for different values of $N$ and $d$: (a)
$d=1$, the black lines (from top to bottom) correspond to $W_N(t)$ vs. $t$ for $N=1,2,\ldots,7$ averaged over
$100000$ realisations, the dashed (red) lines have slope $1/2$; (b) In $d=2$, we plot $a_N\, W_N(t)\,[\ln
t]^N$ vs. $t$ for $N=1,\,2,\,3,\,4$ (from bottom to top) averaged over $100000$ realisations. The prefactor
$a_N$ is chosen so that the curves start at the same point to make the visualisation better. (c) $d=3$, the
black lines correspond to $N=1,\,2,\,3,\,4$ (from top to bottom) averaged over $7000$ realisations.
Analytical results predict $W_N(t)\sim t^{(3-N)/2}$ for $N<3$, $W_N(t)\sim
\ln t$ for $N=3$ and $W_N(t)\sim
{\rm const.}$ for $N=4$. The dashed (red) lines have slopes $1$ ($N=1$), $1/2$ ($N=2$), and $0$ ($N=4$), the
dotted (blue) line is proportional to $\ln t$ ($N=3$).}
\label{fig_num}
\end{figure}

\end{widetext}

In summary, we have presented exact asymptotic results for the mean number of common sites $W_N(t)$ visited
by $N$ independent random walkers in $d$ dimensions. We have shown that as a function of $N$ and $d$ in the
$(N-d)$ plane, there are three distinct regimes for the growth of $W_N(t)$, including in particular, an
anomalous intermediate regime $2< d <d_c(N)=2N/(N-1)$. There are several directions in which our work can be
generalized. For instance, it would be easy to compute the mean number of sites visited exactly by $k$
walkers (out of $N$) up to time $t$ using our result in Eq. (\ref{binom.1}). Here we have restricted only to
the $k=N$ case for simplicity. It would be interesting to consider cases where the walkers have different
step lengths or when they start at different positions~\cite{ITG}. Also, computing the full
distribution of
the number of common sites visited by all walkers remains a challenging open problem.

This work was partly done during M.~V. T.'s several visits at LPTMS, Orsay and
he is very grateful for the warm
hospitality he received there. S.~N. M. acknowledges support by ANR grant 2011-BS04-013-01
WALKMAT and by the
Indo-French Centre for the Promotion of Advanced Research under Project 4604-3.
M.~V. T. acknowledges support
by PALM LABEX ProNet and FP7-PEOPLE-2010-IRSES 269139 DCP-PhysBio grants.

\end{document}